\documentclass[11pt,letterpaper]{JHEP3}
\usepackage{graphics}
\usepackage{epsfig}
\usepackage{amssymb}
\usepackage{stmaryrd}
\usepackage{amsmath}
\usepackage{amsfonts}
\usepackage{mathrsfs}
\usepackage{graphicx}

\def\be{\begin{equation}}
\def\ee{\end{equation}}
\def\ba{\begin{eqnarray}}
\def\ea{\end{eqnarray}}

\newcommand{\scri}{{\mathscr I}}

\title{Modave Lectures on Applied AdS/CFT with Numerics
\footnote{Based on the series of lectures given by Hongbao Zhang at the Eleventh International Modave Summer School on
Mathematical Physics, held in Modave, Belgium, September 2015.}}
\author{Minyong Guo\\
Department of Physics, Beijing Normal University, Beijing, 100875, China
\email{minyongguo@mail.bnu.edu.cn}}

\author{Chao Niu\\
School of Physics and Chemistry, Gwangju Institute of Science and Technology, Gwangju 500-712, Korea\\
\email{chaoniu09@gmail.com}}

\author{Yu Tian\\
School of Physics, University of Chinese Academy of Sciences,
Beijing 100049, China\\
State Key Laboratory of Theoretical Physics, Institute of
Theoretical Physics, Chinese Academy of Sciences, Beijing 100190,
China\\
\email{ytian@ucas.ac.cn}}

\author{Hongbao Zhang\\
Department of Physics, Beijing Normal University, Beijing, 100875, China\\
Theoretische Natuurkunde, Vrije Universiteit Brussel, and The International Solvay Institutes, Pleinlaan 2, B-1050 Brussels, Belgium\\
\email{hzhang@vub.ac.be}}

\abstract{These lecture notes are intended to serve as an introduction to applied AdS/CFT with numerics for an audience of graduate students and others with little background in the subject. The presentation begins with a poor man's review of current status of quantum gravity, where AdS/CFT correspondence is believed to be the well formulated quantum gravity in the anti-de Sitter space. Then we present the basic ingredients in applied AdS/CFT and introduce the relevant numerics for solving differential equations into which the bulk dynamics collapses. To demonstrate how to apply AdS/CFT with numerics, we take the zero temperature holographic superfluid as a concrete example for case study. In passing, we also present some new results, which include the numerical evidence as well as an elegant analytic proof for the equality between the superfluid density and particle density, namely $\rho_s=\rho$, and the saturation to the predicted value $\frac{1}{\sqrt{2}}$ by conformal field theory for the sound speed in the large chemical potential limit.
}

\begin{document}

\section{Introduction}
Different from the other more formal topics in this summer school, the emphasis of these lectures is on the applications of AdS/CFT correspondence and the involved numerical techniques. As theoretical physicists, we generically have a theory, or a paradigm as simple as possible, but the real world is always highly sophisticated. So it is usually not sufficient for us to play only with our analytical techniques when we try to have a better understanding of  the rich world by our beautiful theory. This is how computational physics comes in the lives of theoretical physicists. AdS/CFT correspondence, as an explicit holographic implementation of quantum gravity in anti-de Sitter space, has recently emerged as a powerful tool for one to address some universal behaviors of strongly coupled many body systems, which otherwise would not be amenable to the conventional approaches.  Furthermore, applied AdS/CFT has been entering the era of {\it Computational Holography}, where numerics plays a more and more important role in such ongoing endeavors. Implementing those well developed techniques in {\it Numerical Relativity} is highly desirable but generically required to be geared since AdS has its own difficulties. In the course of attacking these unique difficulties, some new numerical schemes and computational techniques have also been devised. These lectures are intended as a basic introduction to the necessary numerics in applied AdS/CFT in particular for those beginning practitioners in this active field. Hopefully in the end, the readers can appreciate the significance of numerics in connecting AdS/CFT to the real world at least as we do.

The outline of these lecture notes is the following. In the next section, we shall first present a poor man's review of the current status for quantum gravity, where AdS/CFT stands out as the well formulated quantum gravity in anti-de Sitter space. Then we provide a brief introduction to applied AdS/CFT in Section \ref{holography}, which includes what AdS/CFT is, why AdS/CFT is reliable, and how useful AdS/CFT is. In Section \ref{numerics},  we shall present the main numerical methods for solving differential equations, which is supposed to be the central task in applied AdS/CFT. Then we take the zero temperature holographic superfluid as a concrete application of AdS/CFT with numerics in Section \ref{casestudy}, where not only will some relevant concepts  be introduced but also some new results will be presented for the first time. We conclude these lecture notes with some remarks in the end.

\section{Quantum Gravity}
The very theme in physics is to unify a variety of seemingly distinct phenomena by as a few principles as possible, which can help us to build up a sense of safety while being faced up with the unknown world. This may be regarded as another contribution of the unification in physics to our society on top of its various induced technology innovations. With a series of achievements along the road to unification in physics, we now end up with the two distinct entities, namely quantum field theory and general relativity.

As we know, quantum field theory is a powerful framework for us to understand a huge range of phenomena in Nature such as high energy physics and condensed matter physics. Although the underlying philosophies are different, they share quantum field theory as their common language. In high energy physics, the philosophy is reductionism, where the goal is to figure out the UV physics for our effective low energy IR physics. The standard model for particle physics is believed to be an effective low energy theory. To see what really happens at UV, we are required to go beyond the standard model by reaching a higher energy scale. This is the reason why we built LHC in Geneva. This is also the reason why we plan to go to the Great Collider from the Great Wall in China. While in condensed matter physics, the philosophy is emergence. Actually we have a theory of everything for condensed matter physics, namely QED, or the Schrodinger equation for electrons with Coulomb interaction among them. What condensed matter physicists are concerned with is how to engineer various low temperature IR fixed points, namely various phases from such a known UV theory. Such a variety of phases gives rise to a man-made multiverse, which is actually resonant to the landscape suggested by string theory.

On the other hand, general relativity tells us that gravity is geometry. Gravity is different, so subtle is gravity. The very longstanding issue in fundamental physics is trying to reconcile general relativity with quantum field theory. People like to give a name to it, called {\it Quantum Gravity} although we have not fully succeeded along this lane.  Here is a poor man's perspective into the current status of quantum gravity, depending on the asymptotic geometry of spacetime\footnote{This is a poor man's perspective because we shall try our best not to touch upon string theory although it is evident that this perspective is well shaped by string theory in a direct or indirect way throughout these lecture notes.}. The reason is twofold. First, due to the existence of Planck scale $l_p=(\frac{G\hbar}{c^3})^\frac{1}{d-1}$, spacetime is doomed such that one can not define local field operators in a $d+1$ dimensional gravitational theory. Instead, the observables can live only on the boundary of spacetime. Second,  it is the dependence on the asymptopia that embodies the background independence of quantum gravity.

\subsection{De Sitter space: Meta-observables}
\begin{figure}
\center{
\includegraphics[width=4in]{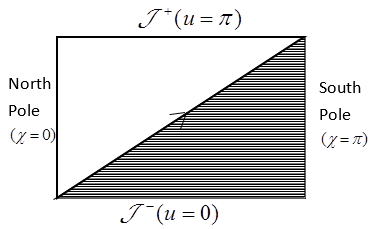}}
\caption{The Penrose diagram for the global de Sitter space, where the planar de Sitter space associated with the observer located at the south pole  is given by the shaded portion.\label{pds}}
\end{figure}

 If the spacetime is asymptotically de Sitter as
 \begin{equation}
 ds^2=-dt^2+l^2\cosh^2\frac{t}{l} d\Omega^2_d,
 \end{equation}
when $t\rightarrow\pm\infty$, then by the coordinate transformation $u=2\tan^{-1}e^\frac{t}{l}$, the metric becomes
\begin{equation}
ds^2=\frac{l^2}{\sin^2 u}(du^2+d\chi^2+\sin^2\chi d\Omega^2_{d-1})
\end{equation}
with $\chi$ the polar angle for the $d$-sphere. We plot the Penrose diagram in Figure \ref{pds} for de Sitter space. Whence both the past and future conformal infinity $\scri^\mp$ are spacelike. As a result, any observer can only detect and influence portion of the whole spacetime. Moreover, any point in $\scri^+$ is causally connected by a null geodesic to its antipodal point in $\scri^-$ for de Sitter.  In view of this, Witten has proposed the meta-observables for quantum gravity in de Sitter space, namely
\begin{equation}
\langle f|i\rangle=\int_{g_i}^{g_f}D g e^{iS[g]}
\end{equation}
with $g_f$ and $g_i$ a set of data specified on $\scri^\pm$ respectively. Then one can construct the Hilbert space $H_i$ at $\scri^-$ for quantum gravity in de Sitter space with the inner product $(j, i)=\langle\Theta j|i\rangle$ by CPT transformation $\Theta$. The Hilbert space $H_f$ at $\scri^+$ can be constructed in a similar fashion. At the perturbative level, the dimension of Hilbert space for quantum gravity in de Sitter is infinite, which is evident from the past-future singularity of the meta-correlation functions at those points connected by the aforementioned null geodesics. But it is suspected that the non-perturbative dimension of Hilbert space is supposed to be finite. This is all one can say with such mata-observables\cite{Witten1}.

However, there are also different but more optimistic perspectives. Among others, inspired by AdS/CFT, Strominger has proposed DS/CFT correspondence. First,  with $\scri^+$ identified as $\scri^-$ by the above null geodesics, the dual CFT lives only on one sphere rather than two spheres. Second, instead of working with the global de Sitter space, DS/CFT correspondence can be naturally formulated in the causal past of any given observer, where the bulk spacetime is the planar de Sitter and the dual CFT lives on $\scri^-$. For details, the readers are referred to Strominger's original paper as well as his Les Houches lectures\cite{Strominger1,SSV}.

 \subsection{Minkowski space: S-Matrix program}
 \begin{figure}
\center{
\includegraphics[width=2.5in]{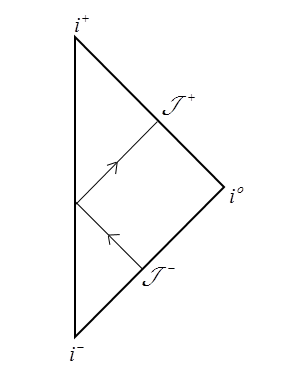}}
\caption{The Penrose diagram for Minkowski space, where massless particles will always emanate from $\scri^-$ and end at $\scri^+$.\label{pmk}}
\end{figure}

The situation is much better if the spacetime is asymptotically flat. As the Penrose diagram for Minkowski space shows in Figure \ref{pmk}, the conformal infinity is lightlike. In this case, the only observable is scattering amplitude, abbreviated as S-Matrix, which connects the out states at $\scri^+$ to the in states at $\scri^-$\footnote{Here we are concerned with the scattering amplitude for massless particles, including gravitons, since they are believed to be more fundamental than massive particles. But nevertheless by taking into account the data at $i^\pm$, the scattering amplitude with massive particles involved  can still be constructed in principle as it should be the case.}. One can claim to have a well defined quantum gravity in asymptotically flat space once a sensible recipe is made for the computation of S-Matrix with gravitons. Actually, inspired by BCFW recursion relation\cite{BCFW}, there has been much progress achieved over the last few years along this direction by the so called S-Matrix program, in which the scattering amplitude is constructed without the local Lagrangian, resonant to the non-locality of quantum gravity\cite{ACK}. Traditionally, S-Matrix is computed by the Feynman diagram techniques, where the Feynman rules come from the local Lagrangian. But the computation becomes more and more complicated when the scattering process involves either more external legs or higher loops. While in the S-Matrix program the recipe for the computation of scattering amplitude, made out of the universal properties of S-Matrix, such as Poincare or BMS symmetry, unitarity and analyticity of S-Matrix, turns out to be far more efficient. It is expected that such an ongoing S-Matrix program will lead us eventually towards a well formulated quantum gravity in asymptotically flat space.

\subsection{Anti-de Sitter space: AdS/CFT correspondence}
\begin{figure}
\center{
\includegraphics[width=2.5in]{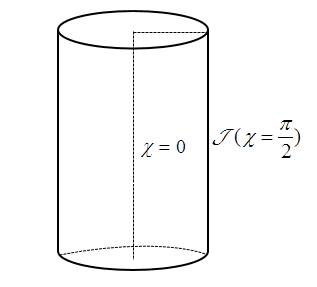}}
\caption{The Penrose diagram for the global anti-de Sitter space, where the conformal infinity $\scri$  itself can be a spacetime on which the dynamics can live.\label{pads}}
\end{figure}

The best situation is for the spacetime which is asymptotically anti-de Sitter as
\begin{equation}
ds^2=\frac{l^2}{\cos^2\chi}(-dt^2+d\chi^2+\sin^2\chi d\Omega^2_{d-1})
\end{equation}
with $\chi\in[0,\frac{\pi}{2})$. As seen from the Penrose diagram for anti-de Sitter space in Figure \ref{pads}, the conformal infinity $\scri$ is timelike in this case, where we can have a well formulated quantum theory for gravity by AdS/CFT correspondence\cite{Maldacena,Witten2,GKP}. Namely the quantum gravity in the bulk AdS$_{d+1}$ can be holographically formulated in terms of CFT$_d$ on the boundary without gravity and vice versa. We shall elaborate on AdS/CFT in the subsequent section. Here we would like to mention one very interesting feature about AdS/CFT, that is to say, generically we have no local Lagrangian for the dual CFT, which echoes the aforementioned S-Matrix program somehow.

\section{Applied AdS/CFT}\label{holography}
\subsection{What AdS/CFT is}
To be a little bit more precise about what AdS/CFT is, let us first recall the very basic object in quantum field theory, namely the generating functional, which is defined as
\begin{equation}
Z_d[J]=\ln[\int D\Psi e^{iS_d[\Psi]+\int d^dx JO}].
\end{equation}
Whence one can obtain the $n$-point correlation function for the operator $O$ by taking the $n$-th functional derivative of the generating functional with respect to the source $J$. For example,
\begin{eqnarray}
&&\langle O(x)\rangle=\frac{\delta Z_d}{\delta J(x)},\\ \label{vev}
&&\langle O(x_1)O(x_2)\rangle=\frac{\delta^2 Z_d}{\delta J(x_1)\delta J(x_2)}=\frac{\delta O(x_1)}{\delta J(x_2)}.
\end{eqnarray}
As we know, we can obtain such a generating functional by perturbative expansion using the Feynman diagram techniques for weakling coupled quantum field theory, but obviously such a perturbation method breaks down when the involved quantum field theory is strongly coupled except one can find its weak dual. AdS/CFT provides us with such a dual for strongly coupled quantum field theory by a classical gravitational theory with one extra dimension. So now let us turn to general relativity, where the basic object is the action given by
\begin{equation} \label{action}
S_{d+1}=\frac{1}{16\pi G}\int d^{d+1}x\sqrt{-g}(R+\frac{d(d-1)}{l^2}+L_{matter})
\end{equation}
for AdS gravity. Here for the present illustration and later usage, we would like to choose the Lagrangian for the matter fields as
\begin{equation}
L_{matter}=\frac{l^2}{Q^2}(-\frac{1}{4}F^{ab}F_{ab}-|D\Phi|^2-m^2|\Phi|^2)
\end{equation}
with $F=dA$, $D=\nabla-iA$ and $Q$ the charge of complex scalar field. The variation of action gives rise to the equations of motion as follows
\begin{eqnarray}
&&G_{ab}-\frac{d(d-1)}{2l^2}g_{ab}=\frac{l^2}{Q^2}[F_{ac}F_b{}^c+2D_a\Phi D_b\Phi-(\frac{1}{4}F_{cd}F^{cd}+|D\Phi|^2+m^2|\Phi|^2)g_{ab}],\\
&&\nabla_aF^{ab}=i(\overline{\Phi}D^b\Phi-\Phi\overline{D^b\Phi}),\\
&&D_aD^a\Phi-m^2\Phi=0.
\end{eqnarray}
Note that the equations of motion are generically second order PDEs. So to extrapolate the bulk solution from the AdS boundary, one is required to specify a pair of boundary conditions for each bulk field at the conformal boundary of AdS, which can be read off from the asymptotical behavior for the bulk fields near the AdS boundary
\begin{eqnarray}\label{conformaldimension}
&&ds^2\rightarrow\frac{l^2}{z^2}[dz^2+(\gamma_{\mu\nu}+t_{\mu\nu}z^d)dx^\mu dx^\nu],\\
&&A_\mu\rightarrow a_\mu+b_\mu z^{d-2},\label{asympt1}\\
&&\Phi\rightarrow\phi_-z^{\Delta_-}+\phi_+z^{\Delta_+}\label{asympt2}
\end{eqnarray}
with $\Delta_\pm=\frac{d}{2}\pm\sqrt{\frac{d^2}{4}+m^2l^2}$\footnote{Here we are working with the axial gauge for the bulk metric and gauge fields, which can always been achieved. In addition, although the mass square is allowed to be negative in the AdS it can not be below the BF bound $-\frac{d^2}{4l^2}$\cite{BF1,BF2}.}. Namely $(\gamma_{\mu\nu}, t_{\mu\nu})$ are the boundary data for the bulk metric field, $(a_\mu,b_\mu)$ for the bulk gauge field, and $(\phi_-,\phi_+)$ for the bulk scalar field. But such pairs usually lead to singular solutions deep into the bulk. To avoid these singular solutions, one can instead specify the only one boundary condition from each pair such as $(\gamma_{\mu\nu},a_\mu,\phi_-)$. We denote these boundary data by $J$, whose justification will be obvious later on. At the same time we also require the regularity of the desired solution in the bulk. In this sense, the regular solution is uniquely determined by the boundary data $J$. Thus the on-shell action from the regular solution will be a functional of $J$.

What AdS/CFT tells us is  that this on-shell action in the bulk can be identified as the generating functional for strongly coupled quantum field theory living on the boundary, i.e.,
\begin{equation}\label{dictionary}
Z_d[J]=S_{d+1}[J],
\end{equation}
where apparently $J$ has a dual meaning, not only serving as the source for the boundary quantum field theory but also being the boundary data for the bulk fields. In particular, $\gamma_{\mu\nu}$ sources the operator for the boundary energy momentum tensor whose expectation value is given by (\ref{vev}) as $t^{\mu\nu}$, $a_\mu$ sources a global $U(1)$ conserved current operator whose expectation value is given as $b^\mu$, and the expectation value for the operator dual to the source $\phi_-$ is given as $\overline{\phi_+}$ up to a possible proportional coefficient. The conformal dimension for these dual operators can be read off from (\ref{conformaldimension}) by making the scaling transformation $(z, x^\mu)\rightarrow(\alpha z, \alpha x^\mu)$ as $d$, $d-1$, and $\Delta_+$ individually.

Here is a caveat on the validity of (\ref{dictionary}). Although such a boundary/bulk duality is believed to hold in more general circumstances, (\ref{dictionary}) works for the large $N$ strongly coupled quantum field theory on the boundary where $N$ and the coupling parameter of the dual quantum field theory are generically proportional to some powers of $\frac{l}{l_p}$ and  $\frac{l}{l_s}$, respectively. In order to capture the $\frac{1}{N}$ correction to the dual quantum field theory by holography, one is required to calculate the one-loop partition function on top of the classical background solution in the bulk. On the other hand, to see the finite coupling effect in the dual quantum field theory by holography, one is required to work with higher derivative gravity theory in the bulk. But in what follows,  for simplicity we shall work exclusively with (\ref{dictionary}) in its applicability regime.

Among others, we would like to conclude this subsection with the three important implications of AdS/CFT. First, a finite temperature quantum field theory at finite chemical potential is dual to a charged black hole in the bulk. Second, the entanglement entropy of the dual quantum field theory can be calculated by holography as the the area of the bulk minimal surface anchored onto the entangling surface\cite{RT,HRT,LM}. Third, the extra bulk dimension represents the renormalization group flow direction for the boundary quantum field theory with AdS boundary as UV, although the renormalization scheme is supposed to be different from the conventional one implemented in quantum field theory\footnote{This implication is sometimes dubbed as $RG=GR$.}.

\subsection{Why AdS/CFT is reliable}
But why AdS/CFT is reliable? In fact, besides its explicit implementations in string theory such as the duality between Type IIB string theory in $AdS_5\times S^5$ and $\mathcal{N}=4$ SYM theory on the four dimensional boundary, where some results can be computed on both sides and turn out to match each other, there exist many hints from the inside of general relativity indicating that gravity is holographic. Here we simply list some of them as follows.
\begin{itemize}
\item Bekenstein-Hawking's black hole entropy formula $S_{BH}=\frac{A}{4l_p^{d-1}}$\cite{Wald}.
\item Brown-Henneaux's asymptotic symmetry analysis for three dimensional gravity\cite{BH}, where the derived central charge $\frac{3l}{2G}$ successfully reproduces the black hole entropy by the Cardy formula for conformal field theory\cite{Strominger2}.
\item Brown-York's surface tensor formulation of quasi local energy and conserved charges\cite{BY}. Once we are brave enough to declare that this surface tensor be not only for the purpose of the bulk gravity but also for a certain system living on the boundary, we shall end up with the long wave limit of AdS/CFT, namely the gravity/fluid correspondence, which has been well tested\cite{HMR}.
\end{itemize}

On the other hand, we can also see how such an extra bulk dimension emerges from quantum field theory perspective. In particular, inspired by Swingle's  seminal work on the connection between the MERA tensor network state for quantum critical systems and AdS space\cite{Swingle}, Qi has recently proposed an exact holographic mapping to generate the bulk Hilbert space of the same dimension from  the boundary Hilbert space\cite{Qi}, which echoes the aforementioned renormalization group flow implication of AdS/CFT.

Keeping all of these in mind, we shall take AdS/CFT as a first principle and explore its various applications in what follows.

\subsection{How useful AdS/CFT is}
As alluded to above, AdS/CFT is naturally suited for us to address strongly coupled dynamics and non-equilibrium processes by mapping the involved hard quantum many body problems to classical few body problems. There are two approaches towards the construction of holographic models. One is called the top-down approach, where the microscopic content of the dual boundary theory is generically known because the construction originates in string theory. The other is called the bottom-up approach, which can be regarded as kind of effective field theory with one extra dimension for the dual boundary theory.

By either approach, we can apply AdS/CFT to QCD as well as the QCD underlying quark-gluon plasma, ending up with AdS/QCD\cite{CLMRW,GKMMN}. On the other hand, taking into account that there are a bunch of strongly coupled systems in condensed matter physics such as high T$_c$ superconductor, liquid Helium, and non-Fermi liquid,
we can also apply AdS/CFT to condensed matter physics, ending up with AdS/CMT\cite{Hartnoll,McGreevy,Herzog,Horowitz,ILM}. Note that the bulk dynamics boils eventually down to a set of differential equations, whose solutions are generically not amenable to an analytic treatment. So one of the central tasks in applied AdS/CFT is  to find the numerical solutions to differential equations. In the next section, we shall provide a basic introduction to the main numerical methods for solving differential equations in applied AdS/CFT.
\section{Numerics for Solving Differential Equations}\label{numerics}
Roughly speaking, there are three numerical schemes to solve differential equations by transforming them into algebraic equations, namely finite different method, finite element method, and spectral method. According to our experience with the numerics in applied AdS/CFT, it is favorable to make a code from scratch for each problem you are faced up with. In particular, the variant of spectral method, namely pseudo-spectral method turns out to be most efficient in solving differential equations along the space direction where Newton-Raphson iteration method is extensively employed if the resultant algebraic equations are non-linear. On the other hand, finite difference method such as Runge-Kutta method is usually used to deal with the dynamical evolution along the time direction. So now we like to elaborate a little bit on Newton-Raphson method, pseudo-spectral method, as well as Runge-Kutta method one by one.
\subsection{Newton-Raphson method}
\begin{figure}
\center{
\includegraphics[width=4in]{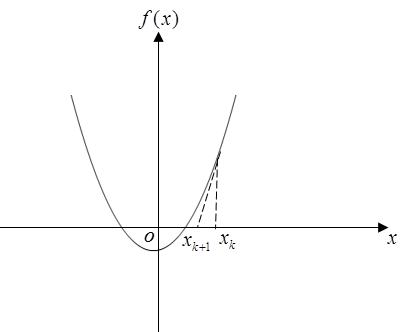}}
\caption{Newton-Raphson iteration map is used to find the rightmost root for a non-linear algebraic equation.\label{NR}}
\end{figure}

To find the desired root for a given non-linear function $f(x)$, we can start with a wisely guessed initial point $x_k$. Then as shown in Figure \ref{NR} by Newton-Raphson iteration map, we hit the next point $x_{k+1}$ as
\begin{equation}
x_{k+1}=x_k-f'(x_k)^{-1}f(x_k),
\end{equation}
which is supposed to be closer to the desired root. By a finite number of iterations, we eventually end up with a good approximation to the desired root. If we are required to find the root for a group of non-linear functions $F(X)$, then the iteration map is given by
\begin{equation}
X_{k+1}=X_k-[(\frac{\partial F}{\partial X})^{-1}F]|_{X_k},
\end{equation}
where the formidable Jacobian can be tamed by Taylor expansion trick since the expansion coefficient of the linear term is simply the Jacobian in Taylor expansion $F(X)=F(X_0)+ \frac{\partial F}{\partial X}|_{X_0}(X-X_0)+\cdot\cdot\cdot$.

\subsection{Pseudo-spectral method}
As we know, we can expand an analytic function in terms of a set of appropriate spectral functions as
\begin{equation}
f(x)=\sum_{n=1}^Nc_nT_n(x)
\end{equation}
with $N$ some truncation number, depending on the numerical accuracy you want to achieve. Then the derivative of this function is given by
\begin{equation}
f'(x)=\sum_{n=1}^Nc_nT_n'(x).
\end{equation}
Whence the derivatives at the collocation points can be obtained from the values of this function at these points by the following differential matrix as
\begin{equation}
f'(x_i)=\sum_jD_{ij}f(x_j),
\end{equation}
where the matrix $D=T'T^{-1}$ with $T_{in}=T_n(x_i)$ and $T'_{in}=T_n'(x_i)$. With this differential matrix, the differential equation in consideration can be massaged into a group of algebraic equations for us to solve the unknown $f(x_i)$ by requiring that both the equation hold at the collocation points and the prescribed boundary conditions be satisfied.

This is the underlying idea for pseudo-spectral method. Among others, we would like to point out the two very advantages of pseudo-spectral method, compared to finite difference method and finite element method. First, one can find the interpolating function for $f(x)$ by the built-in procedure as follows
\begin{equation}
f(x)=\sum_{n,i}T_n(x)T^{-1}_{ni}f(x_i).
\end{equation}
Second, the numerical error decays exponentially with the truncation number $N$ rather than the power law decay followed by the other two methods.

\subsection{Runge-Kutta method}
As mentioned before, we should employ finite difference method to march along the time direction. But before that, we are required to massage the involved differential equation into the following ordinary differential equation
\begin{equation}
\dot{y}=f(y,t),
\end{equation}
which is actually the key step for one to investigate the temporal evolution in applied AdS/CFT. Once this non-trivial step is achieved, then there are a bunch of finite difference schemes available for one to move forward. Among others, here we simply present the classical fourth order Runge-Kutta method as follows
\begin{eqnarray}
&&k_1=f(y_i,t_i),\nonumber\\
&&k_2=f(y_i+\frac{\Delta t}{2}k_1,t_i+\frac{\Delta t}{2}),\nonumber\\
&&k_3=f(y_i+\frac{\Delta t}{2}k_2,t_i+\frac{\Delta t}{2}),\nonumber\\
&&k_4=f(y_i+\Delta tk_3,t_i+\Delta t),\nonumber\\
&&t_{i+1}=t_i+\Delta t, y_{i+1}=y_i+\frac{\Delta t}{6}(k_1+2k_2+2k_3+k_4),
\end{eqnarray}
because it is user friendly and applicable to all the temporal evolution problems we have been considered so far\cite{LTZ,Brussels1,Brussels2,LTZZ,DNTZ,DLTZ}\footnote{It is worthwhile to keep in mind that the accumulated numerical error is of order $O(\Delta t^4)$ for this classical Runge-Kutta method.}.

\section{Holographic Superfluid at Zero Temperature}\label{casestudy}
In this section, we would like to take the zero temperature holographic superfluid as an concrete example to demonstrate how to apply AdS/CFT with numerics. In due course, not only shall we introduce some relevant  concepts, but also present some new results\cite{LNTZ}.

The action for the simplest model of holographic superfluid is just given by (\ref{action}). To make our life easier, we shall work with the probe limit, namely the back reaction of matter fields onto the metric is neglected, which can be achieved by taking the large $Q$ limit. Thus we can put the matter fields on top of the background which is the solution to the vacuum Einstein equation with a negative cosmological constant $\Lambda=-\frac{d(d-1)}{2l^2}$. For simplicity, we shall focus only on the zero temperature holographic superfluid, which can be implemented by choosing the AdS soliton as the bulk geometry\cite{NRT}, i.e.,
\begin{equation}
ds^2=\frac{l^2}{z^2}[-dt^2+d\mathbf{x}^2+\frac{dz^2}{f(z)}+f(z)d\theta^2].
\end{equation}
Here $f(z)=1-(\frac{z}{z_0})^d$ with $z=z_0$ the tip where our geometry caps off and $z=0$ the AdS boundary. To guarantee the smooth geometry at the tip, we are required to impose the periodicity $\frac{4\pi z_0}{3}$ onto the $\theta$ coordinate. The inverse of this periodicity set by $z_0$ is usually interpreted as the confining scale for the dual boundary theory.

In what follows, we will take the units in which $l=1$, $16\pi G Q^2=1$, and $z_0=1$. In addition, we shall focus exclusively on the action of matter fields because the leading $Q^0$ contribution has been frozen by the above fixed background geometry.

\subsection{Variation of action, Boundary terms, and Choice of ensemble}
The variational principle gives rise to the equations of motion if and only if the boundary terms vanish in the variation of action. For our model, the variation of action is given by
\begin{eqnarray}
\delta S&=&\int d^{d+1}x\sqrt{-g}[\nabla_aF^{ab}+i(\Phi\overline{D^b\Phi}-\overline{\Phi}D^b\Phi)]\delta A_b-\int d^dx\sqrt{-h}n_aF^{ab}\delta A_b+ \nonumber\\
&&[(\int d^{d+1}x\sqrt{-g}\overline{(D_aD^a-m^2)\Phi}\delta\Phi-\int d^dx\sqrt{-h}n_a\overline{D^a\Phi}\delta\Phi)+C.C.].
\end{eqnarray}
To make the boundary terms vanish, we can fix $A_b$ and $\Phi$ on the boundary. Fixing $A_b$ amounts to saying that we are working with the grand canonical ensemble. In order to work with the canonical ensemble where $\sqrt{-h}n_aF^{ab}$ is fixed instead, we are required to add the additional boundary term $\int d^3x\sqrt{-h}n_aF^{ab}A_b$ to the action, which is essentially the Legendre transformation. On the other hand, fixing $\phi_-$ gives rise to the standard quantization. We can also have an alternative quantization by fixing $\phi_+$ when $-\frac{d^2}{4}<m^2<-\frac{d^2}{4}+1$\cite{KW}. In what follows, we shall restrict our attention onto the grand canonical ensemble and the standard quantization for the case of $d=3$ and $m^2=-2$, whereby $\Delta_-=1$ and $\Delta_+=2$.

\subsection{Asymptotic expansion, Counter terms, and Holographic renormalization}
What we care about is the on-shell action, which can be shown to have IR divergence generically in the bulk by the asymptotic expansion near the AdS boundary, corresponding to the UV divergence for the dual boundary theory. The procedure to make the on-shell action finite by adding some appropriate counter terms is called holographic renormalization\cite{Skenderis}. For our case, the on-shell action is  given by
\begin{eqnarray}
S_{on-shell}&=&\frac{1}{2}[\int d^4x\sqrt{-g}(\nabla_aF^{ab})A_b-\int d^3x\sqrt{-h}n_aF^{ab}A_b]+ \nonumber\\
&&\frac{1}{2}[(\int d^4x\sqrt{-g}\Phi\overline{(D_aD^a-m^2)\Phi}-\int d^3x\sqrt{-h}n_a\overline{D^a\Phi}\Phi)+C.C.] \nonumber\\
&=&\frac{1}{2}[\int d^4x\sqrt{-g}i(\overline{\Phi}D^b\Phi-\Phi\overline{D^b\Phi})A_b-\int d^3x\sqrt{-h}n_aF^{ab}A_b]- \nonumber\\
&&\frac{1}{2}(\int d^3x\sqrt{-h}n_a\overline{D^a\Phi}\Phi+C.C.).
\end{eqnarray}
By the asymptotic expansion in (\ref{asympt1}) and (\ref{asympt2}), the divergence comes only from the last two boundary terms and can be read off as $\frac{|\phi_-|^2}{z}$\footnote{Note that the outward normal vector is given by $n^a=-z(\frac{\partial}{\partial z})^a$.}. So the holographic renormalization can be readily achieved by adding the boundary term $-\int d^3x\sqrt{-h}|\Phi|^2$ to the original action. Whence we have
\begin{eqnarray}
\langle j^\mu\rangle&=&\frac{\delta S_{ren}}{\delta a_\mu}=b^\mu,\nonumber\\
\langle O\rangle&=&\frac{\delta S_{ren}}{\delta \phi_-}=\overline{\phi_+},
\end{eqnarray}
where $j^\mu$ corresponds to the conserved particle current and the expectation value for the
scalar operator $O$ is interpreted as the condensate order parameter of superfluid. If this scalar operator acquires a nonzero expectation value spontaneously in the situation where the source is turned off, the boundary system is driven into a superfluid phase.

\subsection{Background solution, Free energy, and Phase transition}
\begin{figure}
\begin{center}
\includegraphics[width=7.0cm]{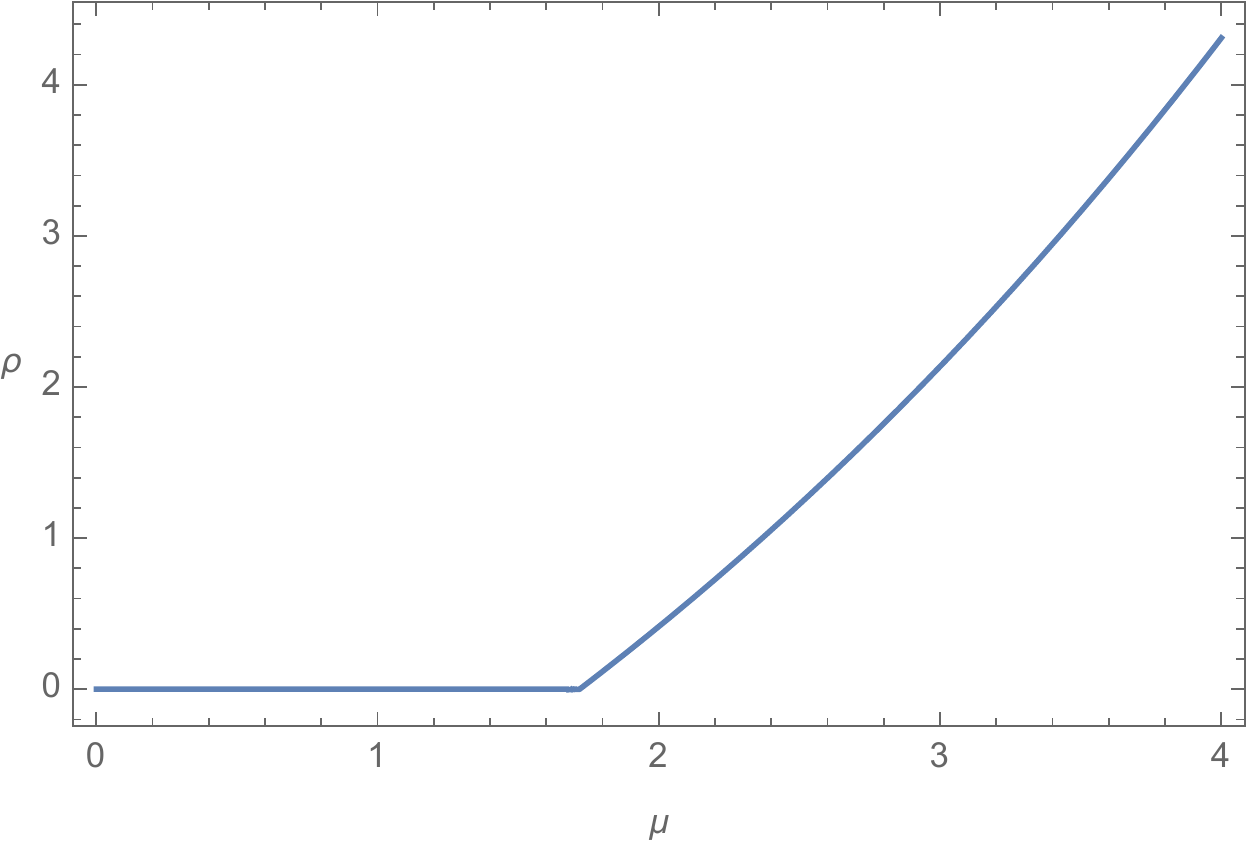}
\includegraphics[width=7.5cm]{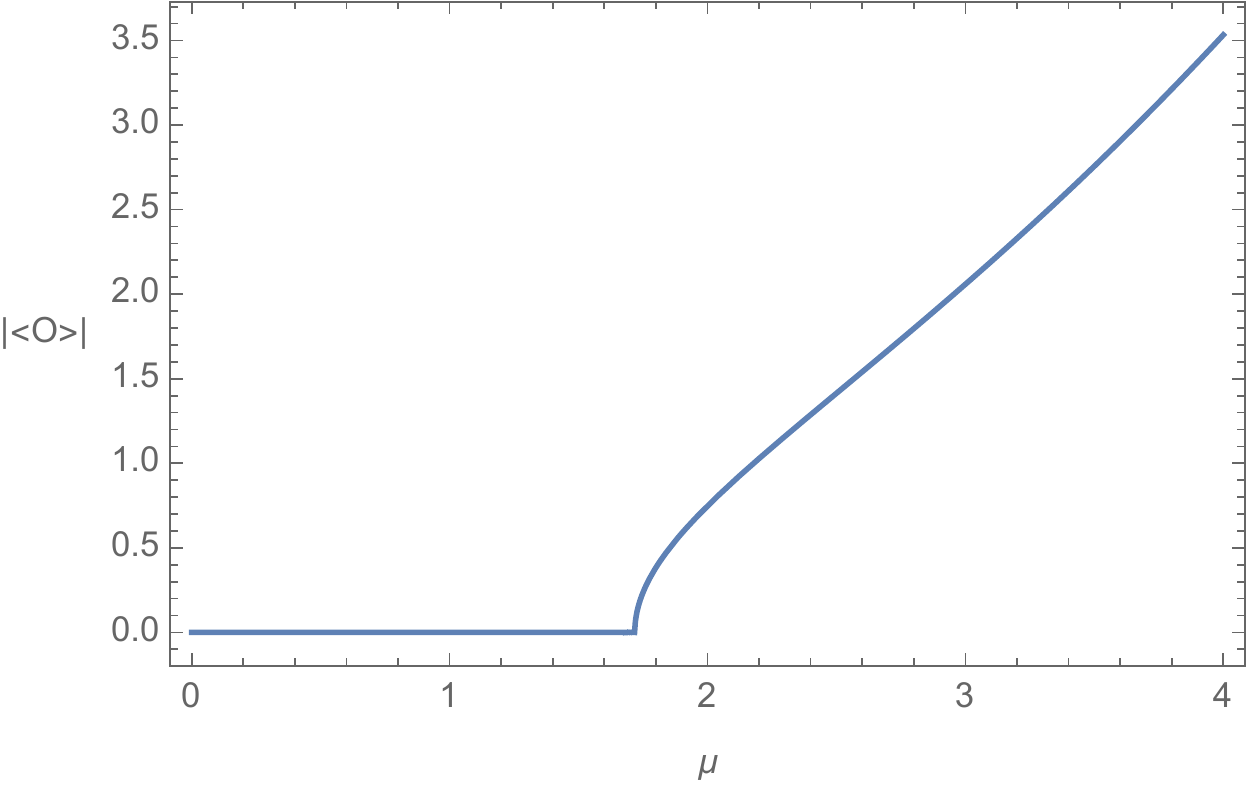}
\end{center}
\caption{The variation of particle density and condensate with respect to the chemical potential, where we see the second order quantum phase transition take place at $\mu_c=1.715$.}
\label{phase}
\end{figure}

With the assumption that  the non-vanishing bulk matter fields $(\Phi=z\phi,A_t,A_x)$ do not depend on the coordinate $\theta$, the equations of motion can be explicitly written as

\begin{eqnarray}
0&=&\partial_t^2\phi+(z+A_x^2-A_t^2+i\partial_xA_x-i\partial_tA_t)\phi+2iA_x\partial_x\phi-2iA_t\partial_t\phi-\partial_x^2\phi\nonumber\\
&&+3z^2\partial_z\phi+(z^3-1)\partial_z^2\phi,\\
0&=&\partial_t^2A_x-\partial_t\partial_xA_t-i(\phi\partial_x\bar{\phi}-\bar{\phi}\partial_x\phi)+2A_x\phi\bar{\phi}+3z^2\partial_zA_x+(z^3-1)\partial_z^2A_x,\\
0&=&(z^3-1)\partial_z^2A_t+3z^2\partial_zA_t-\partial_x^2A_t+\partial_t\partial_xA_x+2\bar{\phi}\phi A_t+i(\bar{\phi}\partial_t\phi-\Psi\partial_t\bar{\phi}),\\
0&=&\partial_t\partial_zA_t+i(\phi\partial_z\bar{\phi}-\bar{\phi}\partial_z\phi)-\partial_z\partial_xA_x, \label{redundant}
\end{eqnarray}
where the third one is the constraint equation and the last one reduces to the conserved equation for the boundary current when evaluated at the AdS boundary, i.e.,
\begin{equation}
\partial_t\rho=-\partial_xj^x.
\end{equation}
To specialize into the homogeneous phase diagram for our holographic model, we further make the following ansatz for our non-vanishing bulk matter fields
\begin{equation}
\phi=\phi(z), A_t=A_t(z).
\end{equation}
Then the equations of motion for the static solution reduce to
\begin{eqnarray}
0&=&3z^2\partial_z\phi+(z^3-1)\partial_z^2\phi+(z-A_t^2)\phi,\\
0&=&2A_t\phi\bar{\phi}+3z^2\partial_zA_t+(z^3-1)\partial_z^2A_t,\\
0&=&\phi\partial_z\bar{\phi}-\bar{\phi}\partial_z\phi,
\end{eqnarray}
where the last equation implies that we can always choose a gauge to make $\phi$ real. It is not hard to see the above equations of motion have a trivial solution
\begin{equation}
\phi=0, A_t=\mu,
\end{equation}
which corresponds to the vacuum phase with zero particle density. On the other hand, to obtain the non-trivial solution dual to the superfluid phase, we are required to resort to pseudo-spectral method.
\begin{figure}
\begin{center}
\includegraphics[width=7.5cm]{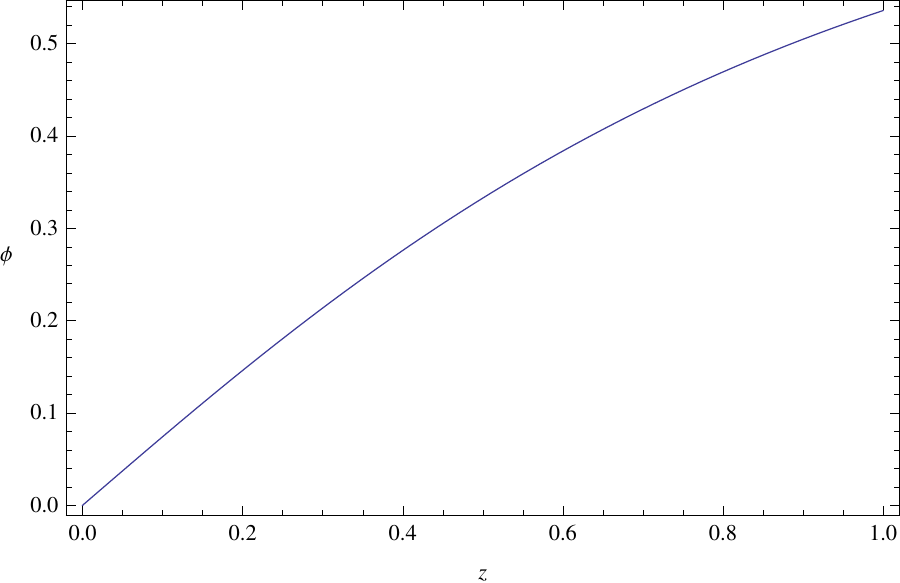}
\includegraphics[width=7.5cm]{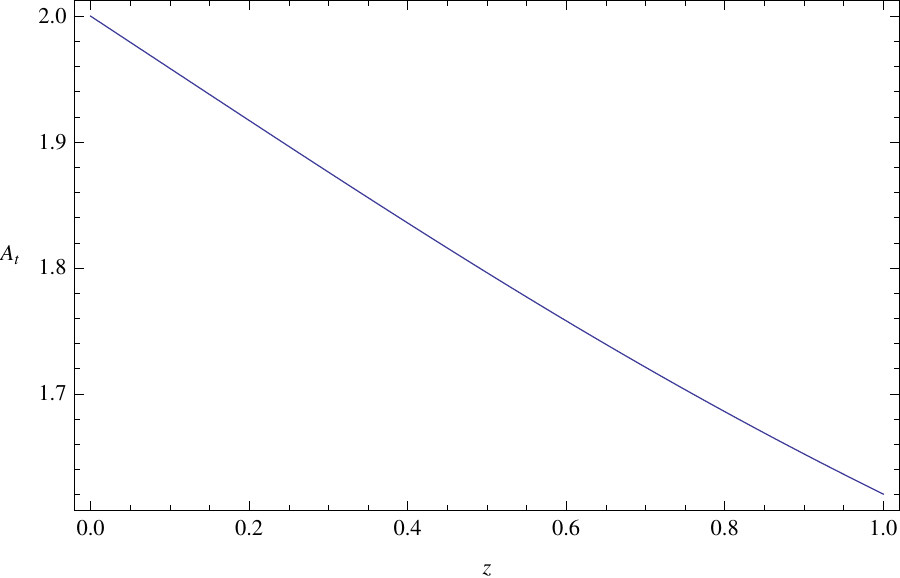}
\end{center}
\caption{The bulk profile for the scalar field and time component of gauge field at the chemical potential $\mu=2$.}
\label{profile}
\end{figure}
As a demonstration, we here plot
the nontrivial profile for $\phi$ and $A_t$ at $\mu=2$ in Figure \ref{profile}. The variation of particle density and condensate with respect to the chemical potential is plotted in Figure \ref{phase}, which indicates that  the phase transition from the vacuum to a superfluid occurs at $\mu_c=1.715$. It is noteworthy that such a phenomenon is reminiscent of the recently observed quantum critical behavior of ultra-cold cesium atoms in an optical lattice across the vacuum to superfluid transition by tuning the chemical potential\cite{ZHTC}. Moreover, the compactified dimension in the AdS soliton background can be naturally identified as the reduced dimension in optical lattices by the very steep harmonic potential as both mechanisms make the effective dimension of the system in consideration reduced in the low energy regime. On the other hand, note that the particle density shows up at the same time as our superfluid condensate, thus it is tempting to claim that this particle density $\rho$ is simply the superfluid density $\rho_s$. This claim is also consistent with the fact that we are working with a zero temperature superfluid where the normal fluid component should disappear. As we will show later on by the linear response theory, this is actually the case.

\begin{figure}
\begin{center}
\includegraphics[width=7.5cm]{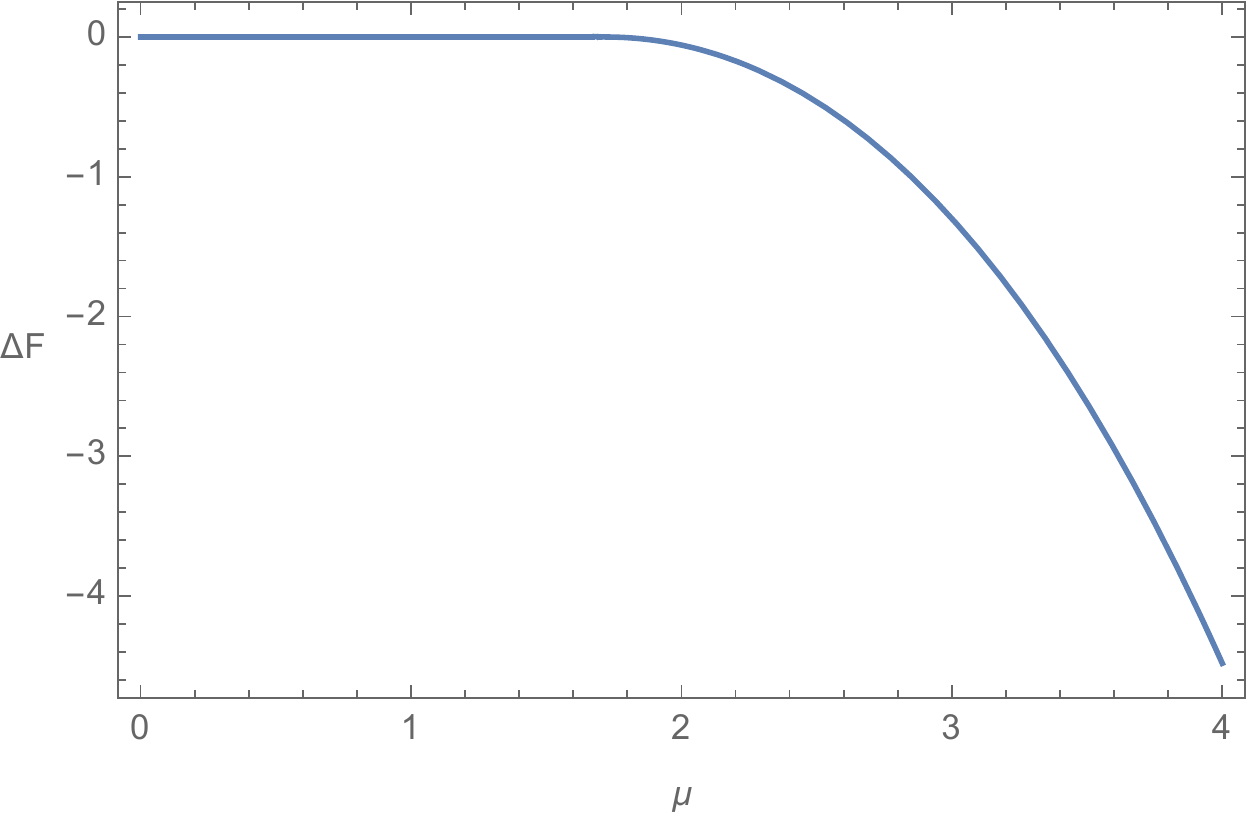}
\end{center}
\caption{The difference of free energy density for the superfluid phase from that for the vacuum phase.}
\label{energy}
\end{figure}

But to make sure that Figure \ref{phase} represents the genuine phase diagram for our holographic model, we are required to check whether the corresponding free energy density is the lowest in the grand canonical ensemble. By holography, the free energy density can be obtained from the renormalized on shell Lagrangian of matter fields as follows\footnote{Here we have used $iS_{Lorentzian}=-S_{Euclidean}$ and $it=\tau$ with $\tau$ the Euclidean time identified as the inverse of temperature.}
\begin{eqnarray}\label{free}
F&=&-\frac{1}{2}[\int dz \sqrt{-g}i(\overline{\Phi}D^b\Phi-\Phi\overline{D^b\Phi})A_b-\sqrt{-h}n_aA_bF^{ab}|_{z=0}]\nonumber\\
&=&-\frac{1}{2}\mu\rho+\int dz(A_t\phi)^2,
\end{eqnarray}
where we have made use of the source free boundary condition for the scalar field at the AdS boundary. As shown in Figure
\ref{energy}, the superfluid phase is the thermodynamically favored one compared to the vacuum phase when the chemical potential is greater than the critical value. So we are done.

\subsection{Linear response theory, Optical conductivity, and Superfluid density}
Now let us set up the linear response theory for the later calculation of the optical conductivity  of our holographic model. To achieve this, we first decompose the field $\phi$ into its real and imaginary parts as
\begin{equation}
\phi=\phi_r+i\phi_i,
\end{equation}
and assume that the perturbation bulk fields take the following form
\begin{equation}
\delta\phi_r=\delta\phi_r(z)e^{-i\omega t+iqx},\delta\phi_i=\delta\phi_i(z)e^{-i\omega t+iqx},\delta A_t=\delta A_t(z)e^{-i\omega t+iqx}, \delta A_x=\delta A_x(z)e^{-i\omega t+iqx},
\end{equation}
since the background solution is static and homogeneous. With this,  the perturbation equations can be simplified as
\begin{eqnarray}
0&=&-\omega^2\delta\phi_r+(z-A_t^2)\delta\phi_r-2i\omega A_t\delta\phi_i+q^2\delta\phi_r+3z^2\partial_z\delta\phi_r+(z^3-1)\partial_z^2\delta\phi_r\nonumber\\
&&-2A_t\phi_r\delta A_t,\\
0&=&-\omega^2\delta\phi_i+(z-A_t^2)\delta\phi_i+2i\omega A_t\delta\phi_r+q^2\delta\phi_i+3z^2\partial_z\delta\phi_i+(z^3-1)\partial_z^2\delta\phi_i
\nonumber\\
&&+i\omega\phi_r\delta A_t+iq\phi_r\delta A_x,\\
0&=&-\omega^2\delta A_x-\omega q\delta A_t+3z^2\partial_z\delta A_x+(z^3-1)\partial_z^2\delta A_x+2\phi_r^2\delta A_x-2iq\phi_r\delta\phi_i,\\
0&=&(z^3-1)\partial_z^2\delta A_t+3z^2\partial_z\delta A_t+q^2\delta A_t+\omega q\delta A_x+2\phi_r^2\delta A_t+4A_t\phi_r\delta\phi_r\nonumber\\
&&+2i\omega\phi_r\delta\phi_i,\\
0&=&-i\omega\partial_z\delta A_t-iq\partial_z\delta A_x-2(\partial_z\phi_r\delta\phi_i-\phi_r\partial_z\delta\phi_i),\label{constraint}
\end{eqnarray}
where we have used $\phi_i=0$ for the background solution.

Note that the gauge transformation
\begin{equation}
A\rightarrow A+\nabla\theta, \phi\rightarrow\phi e^{i\theta}
\end{equation}
with
\begin{equation}
\theta=\frac{1}{i}\lambda e^{-i\omega t+iqx}
\end{equation}
induces a spurious solution to the above perturbation equations as
\begin{equation}
\delta A_t=-\lambda\omega, \delta A_x=\lambda q, \delta\phi=\lambda\phi.
\end{equation}
We can remove such a redundancy by requiring $\delta A_t=0$ at the AdS boundary\footnote{The only exception is the $\omega$ case, which can always be separately managed if necessary.}. In addition, $\delta\phi$ will also be set to zero at the AdS boundary later on. On the other hand, taking into account the fact that the perturbation equation (\ref{constraint}) will be automatically satisfied in the whole bulk once the other perturbations are satisfied\footnote{This result comes from the following two facts.  One is related to Bianchi identity $0=\nabla_av^a=\frac{1}{\sqrt{-g}}\partial_\mu(\sqrt{-g}v^\mu)$ for Maxwell equation, whereby the $z$ component of Maxwell equation satisfies $\partial_z(\frac{v^z}{z^4})=0$ if the rest equations of motion hold. The other is special to our holographic model, in which the readers are encouraged to show that the $z$ component of Maxwell equation turns out to be satisfied automatically at $z=1$ if the rest equations hold there.}, we can forget about (\ref{constraint}) from now on.  That is to say, we can employ the pseudo-spectral method to obtain the desired numerical solution by combining the rest perturbation equations with the aforementioned boundary conditions as well as the other boundary conditions at the AdS boundary, depending on the specific problem we want to solve.

\begin{figure}
\begin{center}
\includegraphics[width=7.5cm]{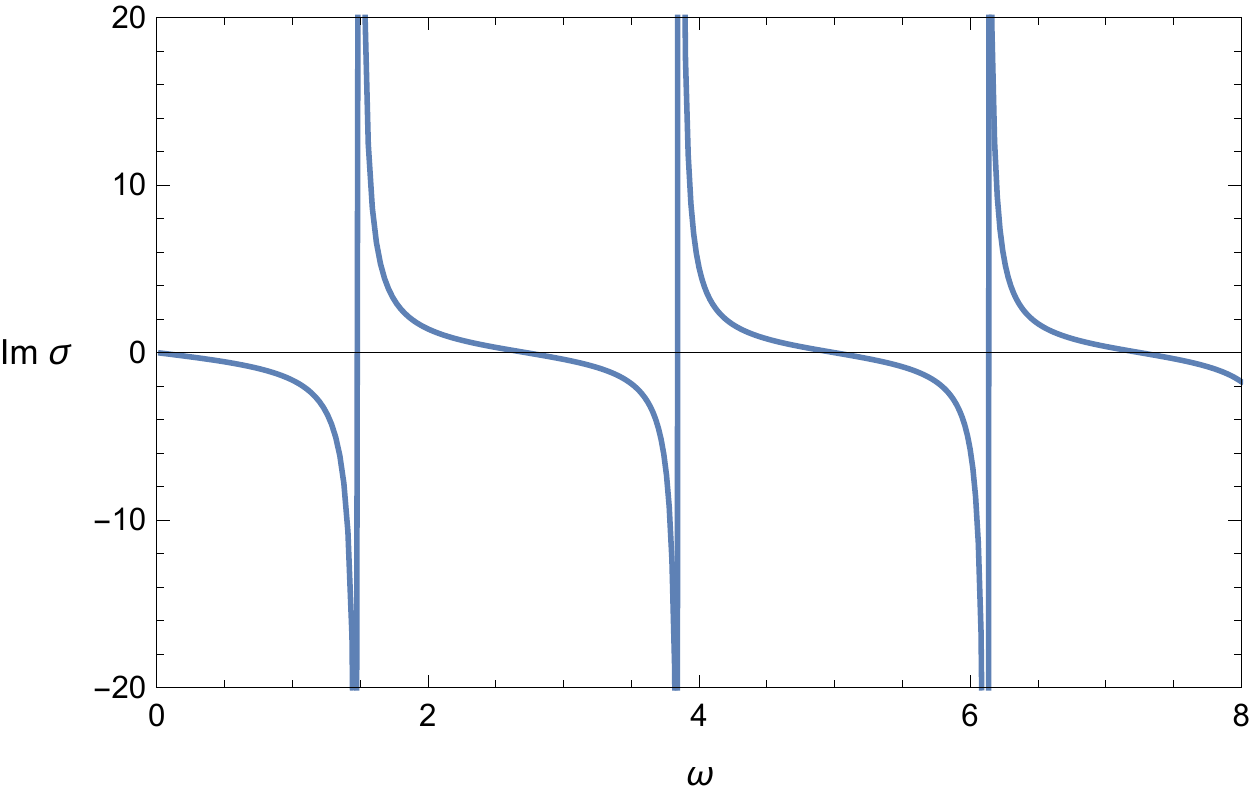}
\includegraphics[width=7.5cm]{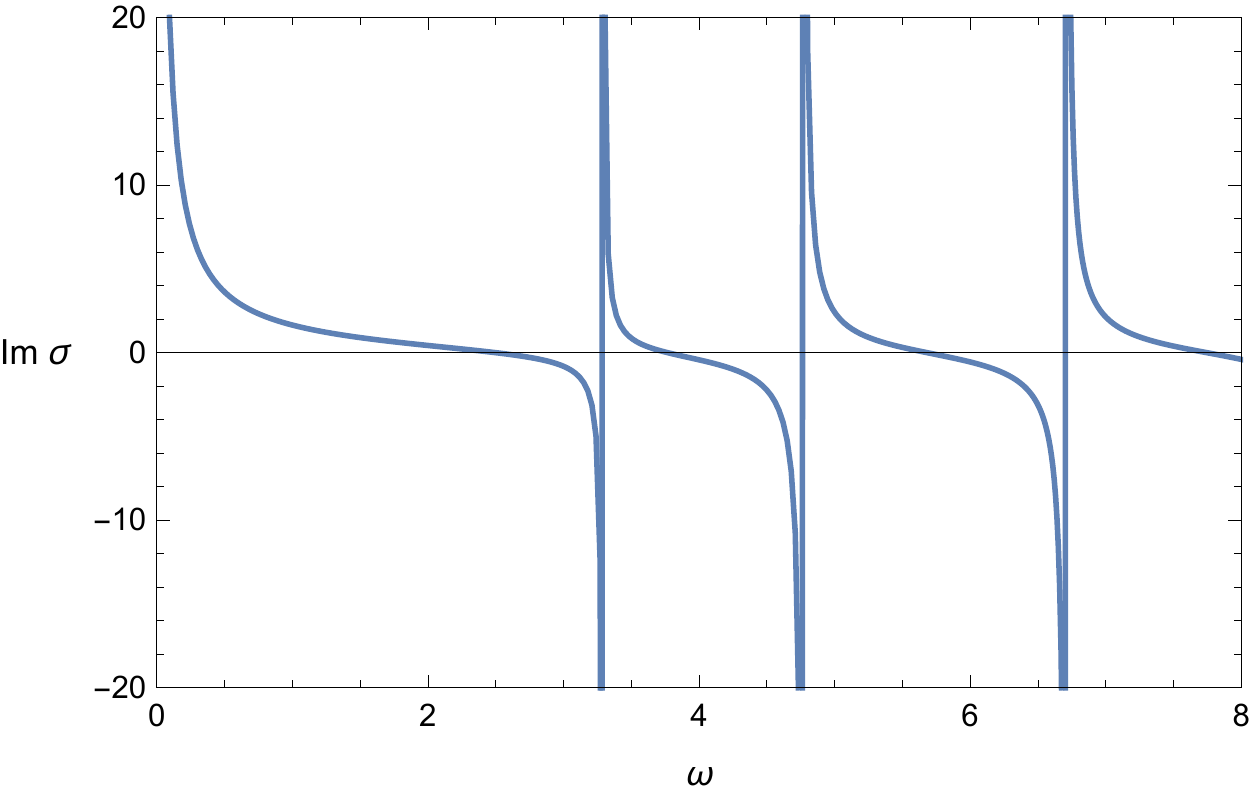}
\end{center}
\caption{The left panel is the imaginary part of optical conductivity for the vacuum phase, and the right panel is for the superfluid phase at  $\mu=6.5$.
}
\label{simple}
\end{figure}

In particular, to calculate the optical conductivity for our holographic model,  we can simply focus on the $q=0$ mode and further impose $\delta A_x=1$
at the AdS boundary.  Then the optical conductivity can be extracted by holography as
\begin{equation}\label{oc}
\sigma(\omega)=\frac{\partial_z\delta A_x|_{z=0}}{i\omega}
\end{equation}
for any positive frequency $\omega$\footnote{Note that $\sigma(-\bar{\omega})=\overline{\sigma(\omega)}$, so we focus only on the positive frequency here.}. According to the perturbation equations,  the whole calculation is much simplified because $\delta A_x$ decouples from the other perturbation bulk fields.
We simply plot the imaginary part of the optical conductivity in Figure \ref{simple} for both vacuum and superfluid phase, because the real part vanishes due to the reality of the perturbation equation and boundary condition for $\delta A_x$. As it should be the case, the DC conductivity vanishes for the vacuum phase, but diverges for the superfluid phase due to the $\frac{1}{\omega}$ behavior of the imaginary part of optical conductivity by the
Krames-Kronig relation
\begin{equation}
\mathbf{Im}[\sigma(\omega)]=\frac{1}{\pi}\mathcal{P}\int_{-\infty}^\infty d\omega'\frac{\mathbf{Re}[\sigma(\omega')]}{\omega-\omega'}.
\end{equation}
Furthermore, according to the hydrodynamical description of superfluid, the superfluid density $\rho_s$ can be obtained by fitting this zero pole as $\frac{\rho_s}{\mu\omega}$\cite{HKS,Yarom1,HY}.
As expected, our numerics shows that the resultant superfluid density is exactly the same as the particle density within our numerical accuracy. The other poles correspond to the gapped normal modes for $\delta A_x$, which we are not interested in since we are focusing on the low energy physics.

Let us come back to the equality between the particle density and superfluid density. Although this numerical result is 100 percent reasonable from the physical perspective, it is highly non-trivial in the sense that the superfluid density comes from the linear response theory while the particle density is a quantity associated with the equilibrium state. So it is better to have an analytic understanding for this remarkable equality. Here we would like to develop an elegant proof for this equality by a boost trick. To this end, we are first required to realize $\rho_s=-\mu\partial_z\delta A_x|_{z=0}$ with $\omega=0$. Such an $\omega=0$ perturbation can actually be implemented by a boost
\begin{equation}
t=\frac{1}{\sqrt{1-v^2}}(t'-vx'), x=\frac{1}{\sqrt{1-v^2}}(x'-vt')
\end{equation}
acting on the superfluid phase. Note that the background metric is invariant under such a boost. As a result, we end up with a new non-trivial solution as follows
\begin{equation}
\phi'=\phi, A_t'=\frac{1}{\sqrt{1-v^2}}A_t, A_x'=-\frac{v}{\sqrt{1-v^2}}A_t.
\end{equation}
We expand this solution up to the linear order in $v$ as
\begin{equation}
\phi'=\phi, A_t'=A_t, A_x'=vA_t,
\end{equation}
which means that the linear perturbation $\delta A_x$ is actually proportional to the background solution $A_t$. So we have $\rho_s=\rho$ immediately.

\begin{figure}
\begin{center}
\includegraphics[width=7.5cm]{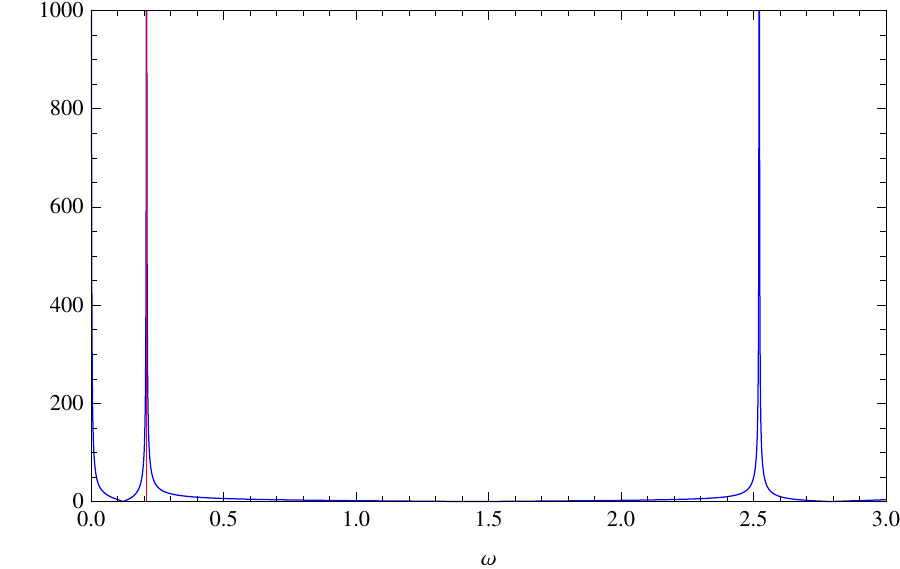}
\end{center}
\caption{ The density plot of $|\frac{det[\mathcal{L}(\omega)]'}{det[\mathcal{L}(\omega)]}|$ with $q=0.3$ for the superfluid phase at $\mu=6.5$. The normal modes can be identified by the peaks, where the red one denotes the hydrodynamic normal mode $\omega_0=0.209$.}
\label{densityplot}
\end{figure}

\begin{figure}
\begin{center}
\includegraphics[width=7.5cm]{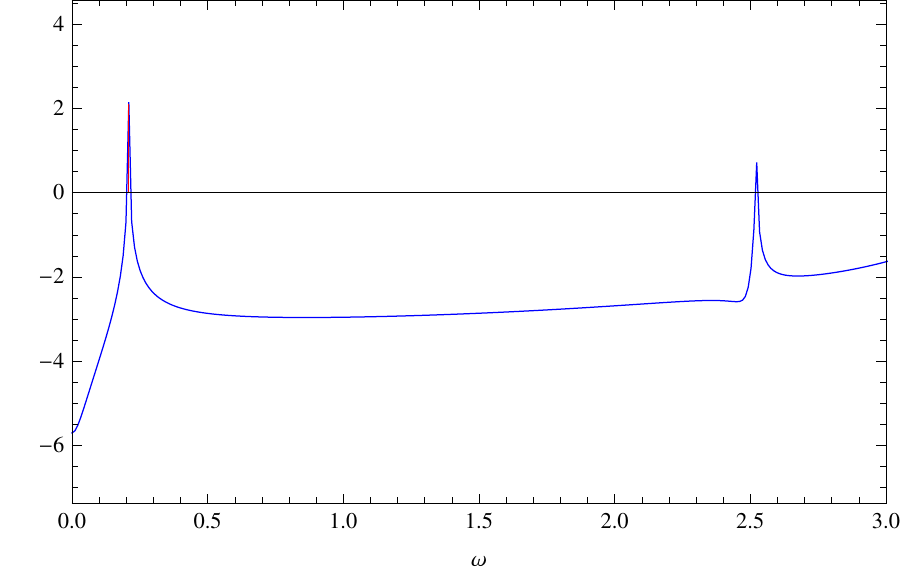}
\end{center}
\caption{ The spectral plot of $\ln{|\delta\hat{\phi}_i(\omega , 1)|}$ with $q=0.3$ for the superfluid phase at $\mu=6.5$, where the initial data are chosen as $\delta\phi_i=z$ with all the other perturbations turned off. The normal modes can be identified by the peaks, whose locations are the same as those by the frequency domain analysis within our numerical accuracy.}
\label{spectralplot}
\end{figure}

\begin{figure}
\begin{center}
\includegraphics[width=7.5cm]{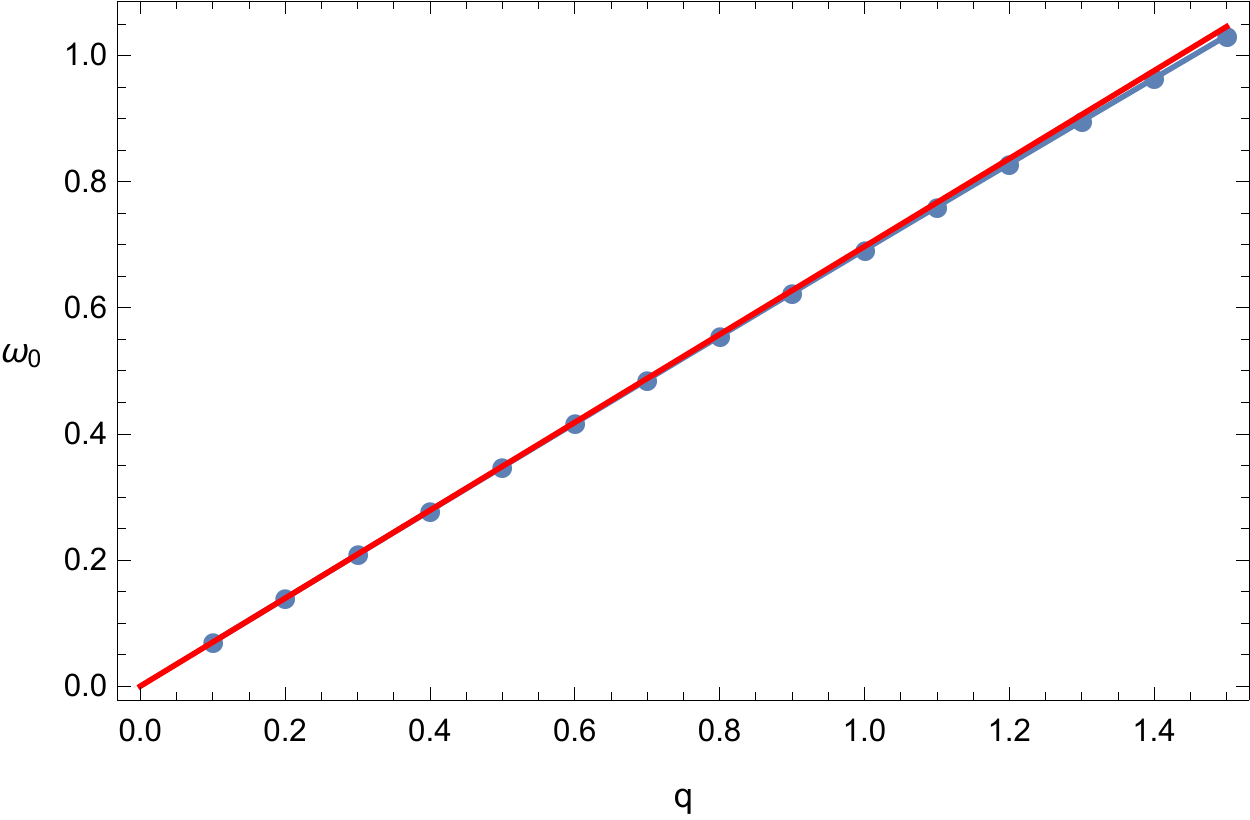}
\end{center}
\caption{ The dispersion relation for the gapless Goldstone mode in the superfluid phase at $\mu=6.5$, where the sound speed $v_s=0.697$ is obtained by fitting the long wave modes with $\omega_0=v_sq$.}
\label{dispersion}
\end{figure}

\begin{figure}
\begin{center}
\includegraphics[width=7.5cm]{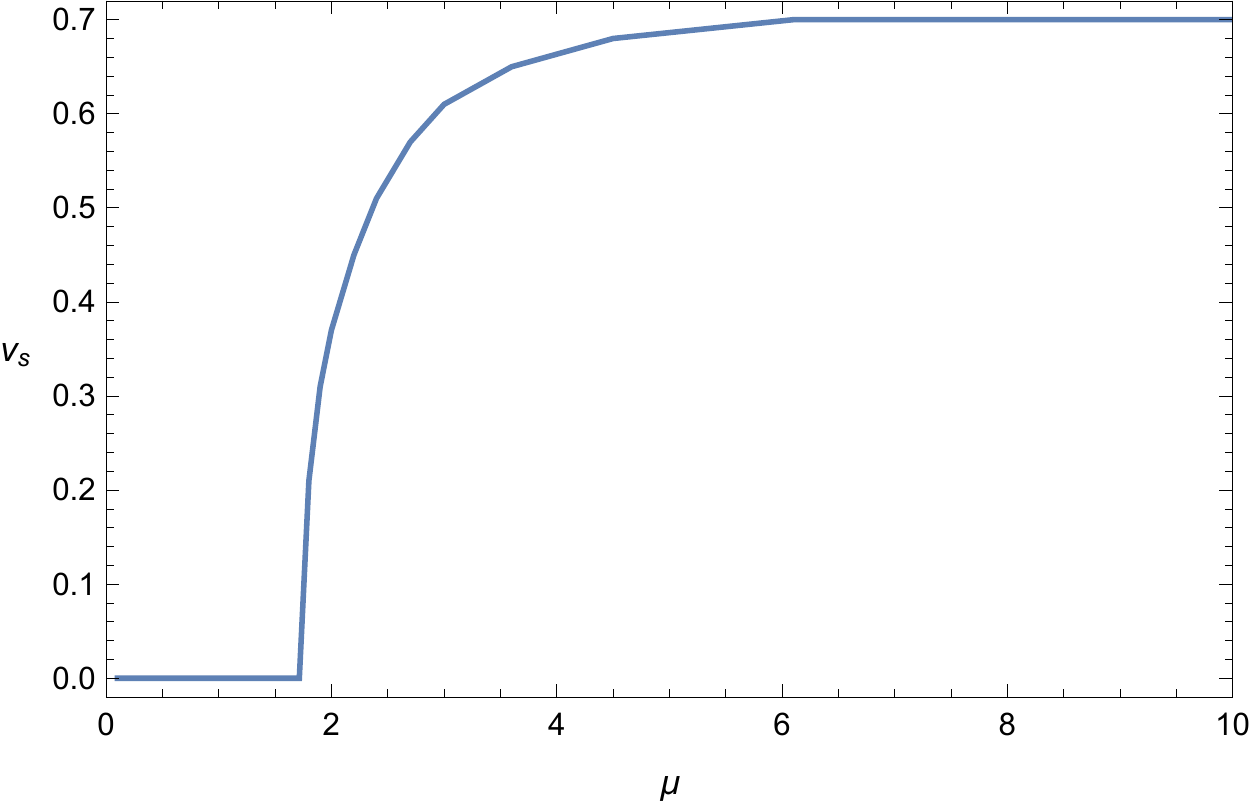}
\end{center}
\caption{ The variation of sound speed with respect to the chemical potential. When the chemical potential is much larger than the confining scale, the conformality is restored and the sound speed approaches the predicted value $\frac{1}{\sqrt{2}}$ by conformal field theory.}
\label{ss}
\end{figure}

\subsection{Time domain analysis, Normal modes, and Sound speed}
In what follows we shall use linear response theory to calculate the speed of sound by focusing solely on the hydrodynamic spectrum of normal modes of the gapless Goldstone from the spontaneous symmetry breaking, which is obviously absent from the vacuum phase.  As such, the perturbation fields are required to have Dirichlet boundary conditions at the AdS boundary. Then we cast the linear perturbation equations and boundary conditions into the form $\mathcal{L}(\omega)u= 0$ with $u$ the perturbation fields evaluated at the grid points by pseudo-spectral method. The normal modes are obtained by the condition $det[\mathcal{L}(\omega)]= 0$, which can be further identified by the density plot  $|\frac{det[\mathcal{L}(\omega)]'}{det[\mathcal{L}(\omega)]}|$ with the prime the derivative with respect to $\omega$.  We demonstrate such a density plot in Figure \ref{densityplot}, where the hydrodynamic mode is simply the closest mode to the origin, marked in red. Besides such a frequency domain analysis of spectrum of normal modes, there is an alternative called time domain analysis, which we would like to elaborate on below. We first cast the equations of motion into the following Hamiltonian formalism
\begin{eqnarray}
\partial_t\phi&=&iA_t\phi+P,\\
\partial_tP&=&iA_tP-(z+A_x^2+i\partial_xA_x)\phi-2iA_x\partial_x\phi+\partial_x^2\phi-3z^2\partial_z\phi+(1-z^3)\partial_z^2\phi,\\
\partial_tA_x&=&\Pi_x+\partial_xA_t,\\
\partial_t\Pi_x&=&i(\phi\partial_x\bar{\phi}-\bar{\phi}\partial_x\phi)-2A_x\phi\bar{\phi}-3z^2\partial_zA_x+(1-z^3)\partial_z^2A_x,\\
0&=&(z^3-1)\partial_z^2A_t+3z^2\partial_zA_t+\partial_x\Pi_x-i(\bar{P}\phi-P\bar{\phi}),\\
\partial_t\partial_zA_t&=&-i(\phi\partial_z\bar{\phi}-\bar{\phi}\partial_z\phi)+\partial_z\partial_xA_x.
\end{eqnarray}
Then with the assumption that  the perturbation bulk fields take the form as $\delta(t,z)e^{iqx}$, the perturbation equations on top of the superfluid phase is given by
\begin{eqnarray}
\partial_t\delta\phi_r&=&-A_t\delta\phi_i+\delta P_r,\\
\partial_t\delta\phi_i&=&\phi_r\delta A_t+A_t\delta\phi_r+\delta P_i,\\
\partial_t\delta P_r&=&A_t\phi_r\delta A_t-A_t\delta P_i-(z+q^2)\delta\phi_r-3z^2\partial_z\delta\phi_r+(1-z^3)\partial_z^2\delta\phi_r,\\
\partial_t\delta P_i&=&-iq\phi_r\delta A_x+A_t\delta P_r-(z+q^2)\delta\phi_i-3z^2\partial_z\delta\phi_i+(1-z^3)\partial_z^2\delta\phi_i,\\
\partial_t\delta A_x&=&\delta\Pi_x+iq\delta A_t,\\
\partial_t\delta\Pi_x&=&2iq\phi_r\delta\phi_i-2\phi_r^2\delta A_x-3z^2\partial_z\delta A_x+(1-z^3)\partial_z^2\delta A_x,\\
0&=&(z^3-1)\partial_z^2\delta A_t+3z^2\partial_z\delta A_t+iq\delta\Pi_x-2\phi_r\delta P_i+2A_t\phi_r\delta\phi_r,\label{unique}\\
\partial_t\partial_z\delta A_t&=&2\partial_z\phi_r\delta\phi_i-2\phi_r\partial_z\delta\phi_i+iq\partial_z\delta A_x.
\end{eqnarray}
As before, using the source free boundary conditions for all the perturbation fields, we can obtain the temporal evolution of the perturbation fields for any given initial data by Runge-Kutta method, where $\delta A_t$ is solved by the constraint equation (\ref{unique}). The normal modes can then be identified by the peaks in the Fourier transformation of the evolving data. We demonstrate such a spectral plot in Figure \ref{spectralplot}. As expected, such a time domain analysis gives rise to the same result for the locations of normal modes as that by the frequency domain analysis.

Then the dispersion relation for the gapless Goldstone can be obtained and plotted in Figure \ref{dispersion}, whereby the sound speed $v_s$ can be obtained by the fitting formula $\omega_0=v_s q$. As shown in Figure \ref{ss}, the sound speed increases with the chemical potential and saturate to the predicted value $\frac{1}{\sqrt{2}}$ by conformal field theory when the chemical potential is much larger than the confining scale\cite{HKS,Yarom1,HY}, which is reasonable since it is believed that the conformality is restored in this limit.

\section{Concluding Remarks}
Like any other unification in physics, AdS/CFT correspondence has proven to be a unique tool for one to address various universal behaviors of near-equilibrium as well as far-from-equilibrium dynamics for a variety of strongly coupled systems, which otherwise would be hard to attack. During such an application, numerical computation has been playing a more and more important role in the sense that not only can numerics leave us with some conjectures to develop an analytic proof and some patterns to have an analytic understanding but also brings us to the regime where the analytic treatment is not available at all.

In these lecture notes, we have touched only upon the very basics for the numerics in applied AdS/CFT. In addition, we work only with the probe limit in the concrete example we make use of to demonstrate how to apply AdS/CFT with numerics. The situation will become a little bit involved when the back reaction is taken into account. Regarding this, the readers are suggested to refer to \cite{DSW} to see how to obtain the stationary inhomogeneous solutions to fully back reacted Einstein equation by Einstein-DeTurck method. On the other hand, the readers are recommended to refer to \cite{CY} to see how to evolve the fully back reacted dynamics, where with a black hole as the initial data it turns out that the Eddington like coordinates are preferred to the Schwarzschild like coordinates.

\begin{acknowledgments}
H.Z. would like to thank the organizers of the Eleventh International Modave Summer School on Mathematical Physics held in Modave, Belgium, September 2015, where the lectures on which these notes are based were given. He is indebted to Nabil Iqbal for his valuable discussions at the summer school.   H.Z. would also like to thank the organizers of 2015 International School on Numerical Relativity and Gravitational Waves held in Daejeon, Korea, July 2015, where these lectures were geared to the audience mainly from general relativity and gravity community. He is grateful to Keun-Young Kim, Kyung Kiu Kim, Miok Park, and Sang-Jin Sin for the enjoyable conversations during the school. H.Z. is also grateful to Ben Craps and Alex Sevrin for the fantastic infrastructure they provide at HEP group of VUB and the very freedom as well as various opportunities they offer to him. M.G. is partially supported by NSFC with Grant Nos.11235003, 11375026 and NCET-12-0054. C.N. is supported by Basic Science Research Program through the National Research Foundation of Korea(NRF) funded by the Ministry of Science, ICT \& Future Planning(NRF- 2014R1A1A1003220) and the 2015 GIST Grant for the FARE Project (Further Advancement of Research and Education at GIST College). Y.T.  is partially supported by NSFC with Grant No.11475179. H.Z. is supported in part by the Belgian Federal
Science Policy Office through the Interuniversity Attraction Pole
P7/37, by FWO-Vlaanderen through the project
G020714N, and by the Vrije Universiteit Brussel through the
Strategic Research Program ``High-Energy Physic''. He is also an individual FWO Fellow supported by 12G3515N.

\end{acknowledgments}

\end{document}